\documentclass{sig-alternate}

\toappear{To appear in the Proceedings of the 5th Grid Computing Environments Workshop (GCE2009),
Portland, Oregon, USA, 2009.}

\usepackage{epstopdf}
\usepackage{subfigure}
\usepackage{color}

\usepackage{pseudocode}  % Formal algorithms
\usepackage{listings}    % Code segments

\usepackage{booktabs}    % Book-style table rules
\usepackage{multirow}    % Row-spanning in tables

\usepackage{stfloats}    % Floats at bottom of page

\usepackage{url}

\begin{document}

\title{{AMP}: A Science-driven Web-based\\Application for the Tera{G}rid}

% === Extensive Author Instructions ===
% Short summary: They want author/affiliation rows with 3 columns,
% and a maximum of 2 rows, for 6 authors total on the front page.

\numberofauthors{3} 

\author{
% 1st. author
\alignauthor
Matthew Woitaszek\\
       \affaddr{National Center for Atmospheric Research}\\
       \affaddr{1850 Table Mesa Drive}\\
       \affaddr{Boulder, CO 80305}\\
       \email{mattheww@ucar.edu}
% 2nd. author
\alignauthor
Travis Metcalfe\\
       \affaddr{National Center for Atmospheric Research}\\
       \affaddr{1850 Table Mesa Drive}\\
       \affaddr{Boulder, CO 80305}\\
       \email{travis@ucar.edu}
% 3rd. author
\alignauthor Ian Shorrock\\
       \affaddr{National Center for Atmospheric Research}\\
       \affaddr{1850 Table Mesa Drive}\\
       \affaddr{Boulder, CO 80305}\\
       \email{ian.shorrock@gmail.com}
}

\maketitle
\begin{abstract}

The Asteroseismic Modeling Portal (AMP) provides a web-based interface
for astronomers to run and view simulations that derive the properties
of Sun-like stars from observations of their pulsation frequencies. In
this paper, we describe the architecture and implementation of AMP,
highlighting the lightweight design principles and tools used to produce
a functional fully-custom web-based science application in less than a
year. Targeted as a TeraGrid science gateway, AMP's architecture and
implementation are intended to simplify its orchestration of TeraGrid
computational resources. AMP's web-based interface was developed as a
traditional standalone database-backed web application using the
Python-based Django web development framework, allowing us to leverage
the Django framework's capabilities while cleanly separating the user
interface development from the grid interface development. We have found
this combination of tools flexible and effective for rapid gateway
development and deployment.

\end{abstract}

% TODO
\category{H.3.5}{Information Storage and Retrieval}{Online Information Services - Web-based services.}

%\terms{}
%\keywords{}

\section{Introduction}

In March 2009, NASA launched the Kepler satellite as part of a mission
to identify potentially habitable Earth-like planets. Kepler detects
planets by observing extrasolar transits--brief dips in observed
brightness as a planet passes between its star and the satellite--that
can be used to identify the size of the planet relative to the size of
the star. However, in order to calculate the absolute size of an
extrasolar planet, the size of the star must also be known.
Asteroseismology can be used to determine the properties of Sun-like
stars from observations of their pulsation frequencies, yielding the
precise absolute size of a distant star and thus the absolute size of
any detected extrasolar planets. The Asteroseismic Modeling Portal
(AMP, \url{http://amp.ucar.edu})
presents a web-based interface to the the MPIKAIA asteroseismology
pipeline \cite{mpikaia} to a broad international community of
researchers, facilitating automated model execution and simplifying data
sharing among research groups.

While the MPIKAIA asteroseismology pipeline itself has been available to
astronomers to download and run on their own resources for several
years, its potential use for processing Kepler data provided compelling
motivation to explore presenting the model as a science gateway. The
most substantial barriers to an astronomer running the model on a local
resource are MPIKAIA's high computational requirements and
straightforward but high-maintenance workflow. Running a single MPIKAIA
simulation requires propagating several independent batches of MPI jobs
and can consume 512 processors for over a week of wall-clock time. More
importantly, the results of these asteroseismology simulations are of
interest to an international community of researchers. Presenting the
model via a science gateway allows researchers without local resources
to run the model, disseminates model results to the community without
repetition, and produces a uniform analysis of asteroseismic data for
many stars of interest.

The straightforward workflow implemented by AMP also provided an
opportunity to develop a new science gateway while exploring a new
architecture, web application framework, and supporting technologies.
One of the first steps when designing a science gateway is to select the
collection of technologies, such as frameworks and toolkits, that will
be used to construct the gateway. As noted by M. Thomas when similarly
evaluating frameworks for science gateway development, gateways can be
constructed using tools that vary greatly in complexity and features,
with the most feature-rich frameworks often introducing substantial
development complexity \cite{thomas-pylons}.
Indeed, many of the prior science gateway projects at the National
Center for Atmospheric Research (NCAR) followed the design
pattern typical of many gateways by using Java to implement complex and
highly-extensible service oriented architectures and web portals.
Most notably divergent from our prior work \cite{gridbgc}, AMP does not
use an application-specific service-oriented architecture and is
not written in Java.

For the design and implementation of AMP, our objective was to create a
web-based science-driven application that peripherally used Grid
technologies to enable the back-end use of supercomputing resources. We
prioritized minimizing development time and complexity while
retaining full creative control of the user interface by selecting the
Django rapid-development web framework and implementing the Grid
functionality with command-line toolkit interfaces. 

Due to AMP's computational requirements, AMP has been designed since its
inception to target TeraGrid resources. Many of the best practices and
procedures for developing and deploying science gateways on the TeraGrid
were proposed coincident with our initial exploration of targeting
TeraGrid as AMP's computational platform. As such, AMP also provides an
example of constructing a new science gateway specifically for TeraGrid
cyberinfrastructure rather than the common case of extending an existing
gateway to utilize TeraGrid. AMP's architecture separates the web-based
user interface and the workflow system performing Grid operations,
isolating interactive users both logically and physically from TeraGrid
operations. We utilized only components common to all TeraGrid resource
providers with the goal of facilitating easy deployment on current
TeraGrid-managed resources without any resource provider assistance.

The remainder of this paper is organized as follows. Section
\ref{sec:background} describes the asteroseismology model workflow and
computational requirements. Section \ref{sec:architecture} and
\ref{sec:implementation} describe the architecture, design, and
implementation of AMP. Section \ref{sec:discussion} discusses our
experiences with AMP's implementation emphasizing the potential
usefulness of the design principles for future gateway projects, and the
paper concludes with continuing and future work.

\section{Background}
\label{sec:background}

The asteroseismology workflow provided by AMP consists of two
components: a forward stellar model and a genetic algorithm (GA) that
invokes the forward model as a subroutine. The forward stellar model is
the Aarhus Stellar Evolution Code (ASTEC) \cite{astec}, a
single-processor code that takes as input five floating-point physical
parameters (mass, metallicity, helium mass fraction, and convective
efficiency) and constructs a model of the star's evolution through a
specified age. The output of the model includes observable data
such as the star's temperature, luminosity, and pulsation frequencies.
In addition to the scalar parameter output, ASTEC produces data that can
be used to produce basic graphical plots describing the star's
characteristics, including a Hertzsprung-Russell diagram showing the
star's temperature and luminosity and an Echelle plot summarizing the
star's oscillation frequencies.

In practice, however, the reverse problem must be solved: ASTEC models a
star with known properties and produces its observable characteristics,
while the real research product requires starting with observations and
identifying the properties of a star that could produce those
observations. In order to derive the properties of distant stars from
observations, ASTEC is coupled with the MPIKAIA parallel GA
\cite{mpikaia} to create an automated stellar processing pipeline
\cite{astec-kepler}. The GA creates a population of candidate stars with
a variety of physical parameters, models each star using ASTEC, and then
evaluates each candidate star for similarity to the observed data. Over
many iterations, the GA converges to identify an optimal candidate star
that has the properties most likely to produce the observed data. The
candidate star is then subjected to a solution detail run that further
refines the star's characteristics at a finer granularity and produces
the final model output.
\\ %NOTE MANUAL CRLF

AMP supports both modes of execution from its web-based user interface:
running the forward model with specific model parameters (a ``direct
model run''), and executing the GA to identify model parameters that
produce observed data (an ``optimization run''). Direct model runs are
trivial to configure and execute: they require five floating-point
parameters as input, take 10-15 minutes to execute on a single
processor, and produce a few kilobytes of output. Optimization runs are
both more complex and computationally intensive. 

\begin{figure}[b]
    \centering
    \epsfig{file=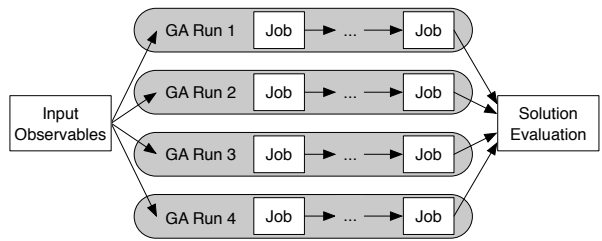, width=3.33in}
    \caption{AMP asteroseismology workflow. }
    \label{fig:workflow}
\end{figure}

%
% Put the table in here so it shows up at the top of the third page.
% I have no idea why the placement is so bizarre.
%
\begin{table*}[t]
    \begin{center}
        \begin{tabular}{ l c c c c c }
            \toprule
            \multirow{2}{*}{System }
                            & Stellar Model        & \multicolumn{4}{c}{Optimization Run (Genetic Algorithm)} \\ \cline{3-6}
                            & Run Time (min)       & Run Time (h) & CPUh    & SUs/CPUh & TeraGrid SUs \\
            \midrule
            NCAR Frost      & 110.0                & 293.3        & 150,187 & 0.558    & 83,804 \\
            NICS Kraken     &  23.6                &  61.9        &  31,723 & 1.623    & 51,486 \\
            TACC Lonestar   &  15.1                &  40.4        &  20,670 & 1.935    & 39,996 \\
            TACC Ranger     &  21.1                &  56.2        &  28,771 & 1.644    & 47,229 \\
            \bottomrule
        \end{tabular}
    \end{center}
    \caption{
        Measured stellar benchmark run time, and estimated optimization
        run time and SU charge, for selected TeraGrid systems. An
        optimization run performs 200 GA iterations and requires about
        160x the model benchmark time to complete, and each GA executes
        four 128-processor jobs.
        }
    \label{table:benchmark}
\end{table*}

The optimization run workflow consists of an ensemble of independent GA
runs, with each run requiring the execution of multiple sequential tasks
(see Figure \ref{fig:workflow}). For each optimization run, multiple
separate GAs are executed and allowed to converge independently. Each GA
(and indeed each task) is started with randomly generated seed
parameters to encourage the GA to explore a wide parameter space, avoid
local minima, and provide confidence in the optimality of the final
result. The GAs can take from hours to days to converge depending on
system performance and the number of iterations requested, so a GA may
not converge in a single task execution within the target
supercomputer's walltime limitations. Thus, each GA run may require
several invocations of the executable to converge to a solution. When
all of the GA runs in the ensemble are complete, the best solution is
evaluated using the forward model to produce detailed output for
presentation and analysis.

In the current configuration for the Kepler data analysis, each
optimization run consists of four GA runs executed in parallel, and each
GA models a population of 126 stars (using 128 processors) for 200
iterations. One interesting artifact of the ASTEC model is that the
execution time varies slightly depending on the target star's
characteristics. During the first few iterations, some stars in the
randomly chosen population may take more time to model than others.
Because the iteration is blocked on the completion of all stars in the
population, the iteration run time is set by the longest-running
component star. However, as the model continues and the population
begins to converge, the model run time for each star also converges and
the time to run each iteration decreases. Thus, the 200 iterations can
be performed in about 160x to 180x of the first iteration's measured
time.

As part of the allocation request for TeraGrid resources, the stellar
model was benchmarked on four TeraGrid platforms (see Table
\ref{table:benchmark}). From the astronomer's perspective, the most
important metric is the predicted optimization run (GA) run time. The
modern Intel and AMD processors in the NICS and TACC resources can
propagate the GA to completion in about 40-60 hours, while the slower
processors in NCAR's Frost system can require over 12 days. When
considering TeraGrid's service unit (SU) charging factors and the model
performance, the TACC systems are most efficient platforms for this
model, but the systems are generally similar in cumulative charging. For
our production deployment, we have targeted the NICS Kraken system due
to its short solution time and support for WS-GRAM. The TACC systems
demonstrated better performance, but the small disk space available on
Lonestar and lack of WS-GRAM on Ranger, combined with the current
allocation oversubscription on those systems, discouraged their use for
this project. For additional computational volume, we continue to
utilize NCAR's Frost system.

\section{Architecture}
\label{sec:architecture}

\begin{figure}[b]
    \centering
    \epsfig{file=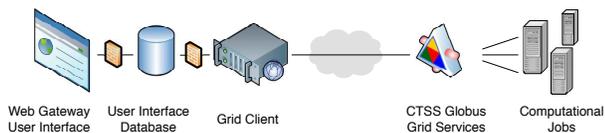,width=3.33in}
    \caption{AMP high-level architecture.}
    \label{fig:architecture}
\end{figure}

The high-level AMP architecture reflects our principal design goals of
supporting rapid development and explicitly targeting TeraGrid
computational resources. The architecture consists of three main
components: the web-based user interface, the ``GridAMP'' workflow
daemon that functions as a grid client, and the remote computational
resources running the model (see Figure \ref{fig:architecture}). The
separation of these three main components is fundamental to the
architecture.

With respect to supporting rapid development, one advantage of the
separation of AMP's functional components is its ability to support
specialized labor. This approach generally decouples the tasks of web
development, back-end Grid software engineering, and the debugging and
maintenance of the science software itself. This is particularly
beneficial because it is much easier to find students to work on
web-related development (e.g., undergraduates) than to find students
that possess a thorough understanding of the intricacies of Grid
infrastructure and middleware (e.g., graduate students with several
years of experience), to say nothing of trying to find students that can
work proficiently (and efficiently) with both. Because the interface and
Grid components are not tightly coupled, they can be easily developed
and maintained by individuals with complimentary skill sets. We have
continued the separation concept through to the science code itself by
running the code in an environment identical to that used by the
astronomy principal investigator and colleagues. Rather than dispatching
software engineers or students to maintain the application, the science
PI occasionally updates the Grid-executed code using {\tt sudo} on the
remote resource personally.

Separating the user interface from the grid-related processing
components also simplifies the administrative responsibilites associated
with using TeraGrid computational resources. In particular, one concern
often associated with science gateways is their use of a shared
credential to submit jobs on behalf of a community of individual gateway
users \cite{tg-gatewaysecuritysummit}. Gateways that utilize TeraGrid
resources are required to maintain user registries and associate every
Grid request with a specific gateway user. In order to provide
end-to-end user accounting for all gateway jobs and to allow resource
providers to disambiguate the real users acting behind community
credentials, TeraGrid has developed and deployed the GridShib SAML
extensions \cite{scavo-gridshib}. However, an underlying risk remains: a
science gateway typically runs a publicly accessible web server and also
must possess the credentials necessary to access many machines on the
TeraGrid.

The AMP architecture addresses this conern by separating users from the
community account credential by placing them on distinct servers. The
user interacts with a web portal located on one publicly-accessible
server, while all back-end processing and remote Grid operations are
performed by the GridAMP daemon on another server. All communication
between the AMP portal and the GridAMP daemon are asynchronously
performed by manipulating a database located on yet another server.
Moreover, the roles and privileges of the public web portal and GridAMP
daemon are strictly managed and controlled. The public web portal is
essentially a database-driven web server without any Grid connectivity
or Grid software. The server hosting the GridAMP daemon is accessible
only to the developers using SSH keys, and only GridFTP is externally
exposed to facilitate data staging via the community account credential.
All input data from users is marshaled through the SQL database.
Incoming user data is parsed by the web server and uploaded to database
tables with strict data type constraints. When required, the input files
are regenerated from the database by the GridAMP daemon and then staged
to TeraGrid systems. It is thus exceptionally difficult to send any data
other than a properly formatted asteroseismology input file to a
TeraGrid resource, and even a full root compromise of the web server
does not provide access to any credentials used for access to any other
system. This architectural feature helps AMP comply at the most
fundamental level with the TeraGrid science gateway security best
practices \cite{tg-gatewaysecurity}.

\section{Implementation}
\label{sec:implementation}

The AMP gateway and the GridAMP daemon are implemented in Python 2.4/2.6
using the Django web development framework \cite{django}. Django's
primary intended use is as a web development platform, but over two
software engineering iterations, we adopted Django as the underlying
framework for both the AMP website and the GridAMP daemon. We were able
to perform two complete cycles of a ``spiral-model'' software
engineering process in about one year, completely re-implementing the
entire website and processing daemon about 6 months after the initial
prototyping commenced.

In our first development prototyping cycle, we perhaps took the
separation of components concept too far, as we used Django to implement
the website but implemented the GridAMP daemon in Python using
manually-coded SQL database calls. This made sense at the time: although
Django provides a full-featured object-relational model (ORM)
independent of its web server-related features, we were skeptical that
the ORM would be sufficiently robust to fulfill our requirements. For
example, we demand direct and explicit control of the database schema
and wanted to use database permissions to carefully control access to
database tables on a per-user basis. Even the idea of allowing a ORM
system to create tables based on Python object definitions seemed
irreconcilable with production-quality science gateway implementation.
Over the first six months of development, however, it became clear that
this was not the case -- the Django ORM was more powerful and flexible
than we imagined could be possible. We were able to easily redefine our
prior manually-specified database schema entirely using Django with
perfect table/field/type correspondence, including our desired
permissions scheme, all from within Django's ORM. Moreover, the database
schema could be reconstructed on demand--including sample data--in test
databases when required for development work. The ORM also worked from
standalone programs outside of Django's web serving infrastructure.

Thus, the usefulness of the Django ``don't repeat yourself'' philosophy
quickly became apparent and immediately applicable to AMP. While the
service separation philosophy can be taken to an extreme -- we could
have even switched languages between the web server and the GridAMP
daemon -- maintaining two separate codebases quickly became a mundane
waste of time. We therefore maintained the operational separation of the
web site and GridAMP daemon but unified the framework for both
components. The entire project now uses a single code base to define and
manipulate shared data structures across multiple servers.

\subsection{Common Components}

Software written with the Django framework is organized into
``projects'' and ``applications''. A project basically represents a
website and consists of a common configuration and a collection of
installed possibly independent applications. Applications are written
using the typical model-view-controller design pattern, better described
as model-template-view using Django's terminology. Models use the ORM to
abstract database access behind Python objects while providing the
opportunity to add custom functionality. When a HTTP request is
received, the request is dispatched to the appropriate Python subroutine
(a ``view'') to perform necessary processing. View routines then usually
conclude by rendering final output to the user via Django's template
engine.

For AMP, we implemented most of the science gateway functionality in a
single core application consisting of ORM models and support routines.
For example, the catalog of stars, their identifiers, the simulations,
and the constituent supercomputer jobs are all stored in this core
application. This effectively makes the most important components of AMP
first-class global objects when imported properly. The web interface is
then constructed of additional applications that refer to the core
application as required. Only this core application's models are shared
between the website and the GridAMP daemon.

For both the web server and the GridAMP daemon, we also adopted Django's
built-in authentication ``auth'' framework. The authentication framework
provides basic website user management functionality including common
user-initiated account manipulation activities. We extended the Django
authentication framework to support additional information required by
AMP and TeraGrid, such as data provenance and user authentication
metadata.

An additional benefit of using the Django ORM and authentiation
framework is that Django's built-in development server provides an
administrative interface that can manipulate ORM objects including those
created by the authentication framework. The interface is also easily
modified to support custom requirements. Thus, administrative tasks such
as approving users or adjusting back-end parameters (like allocations
and the authorization for a user to submit to a machine using a
particular allocation) can easily be manipulated from a graphical
interface without custom development. The interface is available to
developers running the Django development server with appropriate
database connectivity, so the administrative functionality is not even
possible from any publicly accessible web servers.

%
% Again, this is up a bit before the target page so it lands in
% the proper spot.
%

\begin{figure*}[b]
\lstset{language=Python,
        caption=Example GridAMP workflow definition,
        label=workflow,
        showspaces=false,
        showstringspaces=false
}
\begin{lstlisting}[frame=single]
self.workflow = {
    `QUEUED':  ( [self.check_queued_sim, self.submit_prejob],                 `PREJOB'),
    `PREJOB':  ( [self.check_prejob, self.submit_workjob],                    `RUNNING'),
    `RUNNING': ( [self.check_workjob, self.submit_postjob],                   `POSTJOB'),
    `POSTJOB': ( [self.check_postjob, self.postprocess, self.submit_cleanup], `CLEANUP'),
    `CLEANUP': ( [self.check_cleanup, self.close_simulation],                 `DONE')
}
\end{lstlisting}
\end{figure*}

\subsection{User Interface}

In addition to the shared Django application that contains the core AMP
models, we wrote separate Django applications to implement independent
portions of the website functionality. One application allows users to
browse and search star catalogs, one allows users to view completed
simulation results, and another facilitates simulation submission. These
applications don't contain models so they are useful only within the
context of a Django project containing the core AMP application, but the
distinction provided a logical separation of site components.

We also wrote additional standalone Django applications containing
potentially reusable code. For example, we wished to use a CAPTCHA
to reduce the possibility of automated bots requesting AMP accounts. Due
to our accessibility requirements, using a typical image-only CAPTCHA
was problematic, so we decided to write our own. Our general purpose
question/answer CAPTCHA presents a series of questions with optional
links to answers. For AMP, users are asked to enter the HD catalog
numbers of popular stars, such as ``What is the HD number for Alpha
Centauri?'' For astronomers that can't remember, we present a link to
the page containing the answer.
With this, only one real estate agent turned fashion supermodel has
requested the ability to submit AMP jobs.

AMP's web interface is quite typical for current database-driven
websites in that it combines static and dynamic web technologies to
provide its user experience. AMP uses AJAX-based ``Web 2.0'' techniques
to simplify the user experience where possible, but the site is fully
functional without these JavaScript enhancements. For example, the
process of searching for a star uses AJAX to suggest stars with results
or in the Kepler catalog. If no stars are in AMP's catalog, the search
is passed to the SIMBAD \cite{simbad} astronomical database and the
target, if found, is added to the local catalog. Finally, AMP uses
Django's SSL authentication and session management support to ensure
that all activities performed by registered users is encrypted.

\subsection{Grid Execution}

To simplify the deployment of the AMP model on TeraGrid systems, we
constructed a workflow that utilizes only basic components provided by
the Coordinated TeraGrid Software and Services (CTSS) software stack
\cite{ctss}. Rather than deploying a SOA with services that encapsulate
the models as we have done in the past for other projects, the GridAMP
daemon directly formulates and submits GRAM execution requests and
GridFTP file transfers. Thus, the model can be deployed on a TeraGrid
resource as soon as the community account has been authorized and no
special resource provider dispensations (e.g., custom Globus containers
or separate service hosting platforms) are required.

The remote resource execution environment for each AMP job is
initialized and finalized using shell scripts invoked by GRAM using the
fork job service. The pre-job stage creates a new empty copy of the
model runtime directory structure and prepopulates the tree with static
input files. The model is then run using GRAM through the scheduler
interface with each model invocation staging in the small input data
text file and staging out its restart progress file. The post-job stage
uses {\tt tar} to consolidate output and log files into a single file
for transfer back to the GridAMP daemon and eventual delivery to the
user via the website. A final cleanup stage ensures that the execution
environment has been removed.

\subsection{GridAMP Workflow Daemon}

The GridAMP daemon manages the workflow of AMP simulations on remote
grid resources. It reads simulation information from the centralized
database, performs the necessary grid client actions, and updates the
database accordingly. The AMP website and the GridAMP daemon thus
interact asynchronously through the centralized database.

We wrote a custom Python module to handle the grid client functionality
via calls to the Globus command-line interfaces. The module supports
generating derivative proxy certificates with GridSHIB SAML extensions,
GridFTP, and GRAM. The primary reasons for using our own library were
that we already had such functionality in-house and our familiarity with
our grid support module made it seem simpler and more robust than using
third-party solutions. The most important operational benefit for
wrapping command line clients is that it provides excellent support for
troubleshooting. The daemon produces logs that clearly highlight
warnings and errors with the relevant command lines displayed for
failure cases. To troubleshoot, a developer needs only to open a new
console on the GridAMP server and copy-paste the line at the shell
prompt to retry the failed action. The Grid operations are not hidden
behind complex object models but are transparent so that problems can be
investigated and corrected quickly and easily.

Due to AMP's straightforward processing requirements, we also wrote our
own workflow management daemon. The workflow is represented as a list of
stages with function pointers that must return to proceed to the next
state (see Listing \ref{workflow}). If the job is in a particular state,
all of the functions in the subsequent list are called. If all return
True, then the job is set to the indicated next state. In practice, the
first function usually checks to see if the prior state has completed,
and the last function propagates the job to the next state. This simple
encoding can represent arbitrary trees of execution, but for AMP the
processing is merely linear. The only coding cleverness is the use of
inheritance to support AMP's two job types with a single base class
implementing all of the routine functionality. Job queuing, stage-in,
and stage-out are all handled by the base class. Only the functions that
generate the GRAM job definitions and perform model postprocessing are
implemented in the derived classes. Thus, the derived classes are very
small and contain only model-specific execution and postprocessing code.

Workflow state management and job status tracking are integrated with
AMP's data model as implemented using the Django ORM and stored in the
centralized database. We utilized a two-level approach to workflow
status management, integrating the simulation status in the
application-specific data models while maintaining constituent grid job
status in a more generic fashion. To manage the workflow, the daemon
first polls the status of each grid job and updates the job records
accordingly. This process is identical for all grid jobs regardless of
purpose (pre-job, post-job, or simulation) or execution method (fork or
queue), and no special callbacks or processing are performed as part of
the grid job status update procedure. Once the grid job status has been
updated, the workflow management code simply retrieves the last-known
status of the appropriate job and waits or proceeds accordingly. One
advantage to this approach is that simulation status is integrated at
the highest level of the application-specific data model so the user
interface does not need to analyze the state of many individual grid
jobs to determine the current state of a simulation.

As part of the workflow management process, the GridAMP daemon also
handles failures and provides user status notifications. Our error
management philosophy completely isolates gateway users from the jargon
of grid-related failures and transients. Users are not notified of
events that they may not understand and are definitely not capable of
correcting. Unless the asteroseismology model fails, the simulation will
be completed and returned to the user. Users may opt to receive an
e-mail when their simulation completes or to receive e-mails at each
state transition.

The GridAMP daemon distinguishes between anticipated transients, model
processing failures, and its own failures. Anticipated transients, such
as remote systems suddenly becoming unreachable for GRAM or GridFTP
requests, are handled silently: administrators are notified, the job's
status display is supplemented with a plain-text message describing the
situation, and the processing is retried automatically without user or
administrator intervention. Model failures, such as the absence of a
mandatory output file or the failure of a result line to parse
correctly, generally require gateway administrator intervention and
occasionally escalate to the science investigators for model development
work. In the event of a model failure, the simulation is moved to a
special ``hold'' state and both the user and administrator are notified.
The gateway administrators can then debug the problem and retry the
failed processing steps interactively. Once the problem has been
resolved, the workflow resumes automatically. Finally, failures of the
GridAMP daemon itself are monitored externally and immediately brought
to the attention of the gateway administrators. 

%The GridAMP daemon is quite small. The grid support module contains
%only 582 non-comment lines. The workflow manager contains 739 lines,
%with the child classes for direct simulations and optimization runs
%containing 295 and 809 lines, respectively. Most of the code performs
%output processing; that is, reading the output files from the model
%and extracting relevant parameters. An additional 478 lines of code
%are used to generate the plots. If we had used a commodity grid
%control framework, only the grid library would have been replaced.
%If we had used a commodity workflow manager, only the base class of
%the workflow manager 

\section{Discussion}
\label{sec:discussion}

Perhaps the most fundamental characteristic of AMP is its posture as a
{\em grid-enabled} science gateway. When considering our earlier grid
gateway projects and a small set of existing grid gateway frameworks, we
realized that we did not really want to build a ``grid gateway'' in the
sense suggested by these projects and frameworks. Rather, we wanted a
science-driven web-based application focused on delivering the required
functionality to our user community that happened to use grid resources
and technology to perform some of its computationally intensive
processing. To that end, AMP completely hides many aspects of its
grid nature from users. As most astronomers are familiar with
high-performance computing, concepts such as simulations, computational
jobs, allocations, and supercomputers remain visible terminology, but
the word ``certificate'' is not even mentioned anywhere on the site.

Our ability to decouple AMP's front-end and back-end components was
enabled by AMP's straightforward workflow and lengthy job turnaround
time. We recognize that the luxury of asynchronous coupling is not
afforded by many science gateways that facilitate interactive analysis
and visualizations. The decoupled asynchronous processing is appropriate
for AMP's jobs, simplified the implementation, and facilitates
operational debugging.

While workflow management is well understood and a variety of robust
technologies are available to automate workflows \cite{dagman}, it was
indeed quite simple to implement a small-scale custom workflow manager
for AMP. In fact, if GRAM ever supports executing pre-job and post-job
scripts using the fork service as part of a queued job specification,
half of AMP's functionality could be implemented using a single GRAM job
submission! For the optimization runs, the most complex portion of the
workflow is downloading and interpreting partial result files, which
requires custom implementation regardless of the workflow management
paradigm. By writing our own simple workflow management daemon, we have
retained a single application-defined representation of all state. The
Django models used by the website are used for execution management by
the GridAMP daemon. This avoids the need to deploy and query middleware
to run grid jobs and provides the transparent end-to-end debugging
capability that is useful when things go wrong.

We are particularly impressed with the Python-based Django web
development framework. For our purposes, Django seemed to perfectly
balance framework features and customization, supporting the rapid
development web sites without being a content management system. The
programming methodology was intuitive, suggesting but not enforcing a
model-view-controller design pattern. The Django framework was useful
even for the non-web portions of the project. The self-contained
development environment was easy to install and facilitated quick
prototyping and debugging. When combined with the Apache web server, the
framework was robust enough to function as a production system.

Our use of AJAX and Web 2.0 technologies has been limited to cases where
it is clearly beneficial to our user community. For example, the star
search functionality suggests stars that are in the Kepler catalog and
stars that have results as soon as a user types enough of a catalog
identifier to disambiguate possible targets. Given the long job
turnaround time, however, opportunities to make the website appear more
dynamic are limited. We could do many cool tricks with AJAX and social
networking, and it was very tempting to allow astronomers to ``share a
star'' via Facebook or send simulation progress updates using Twitter.
More pragmatically, we are currently working on using RSS feeds to allow
astronomers to subscribe to stars of interest and adding dynamic links
to astronomical catalogs and visualization services such as SIMBAD and
Google Sky.

%The merchandising options via the International Star Registry...

%
%The Django framework has also provided an efficient development process
%and supports production system administration. DjangoÕs built-in
%development server allows developers to test any component of the AMP
%portal from a workstation without installing or configuring a standalone
%web server. In addition, the Django administrative interface allows easy
%manipulation of all underlying database tables without custom software
%development. Routine gateway resource and user management is performed
%using this built-in administrative interface from trusted internal
%systems.

Although AMP was designed as a custom solution for a specific model and
workflow, we believe that some AMP components may be a useful foundation
for future similar grid gateway development. Of course, the AMP user
interface is completely custom, but Django facilitates rapid web
development in its own right. The core AMP models that represent jobs
and the base classes of the workflow manager are potentially
generic enough to support other applications and workflows with
minimal changes. Although we have not done so, it would not be
particularly difficult to isolate the common job management
functionality from the models such that it could be added to new models
as desired. The GridAMP daemon already supports this abstraction, as the
workflow manager base class itself contains only grid code and all
application-specific logic is contained in the workflow-specific derived
classes. This level of abstraction would have to be similarly introduced
to the data models by using complementary table schemas or inheritance
to make a model represent grid jobs using a mechanism other than copying
and pasting certain fields into the model definition. In this more
generic approach, models would be defined only with application-specific
job fields (such as input and results) with the job management fields
provided externally. Thus, while AMP and its underlying components are
clearly not a framework from which new gateways may easily be
constructed, AMP demonstrates how rapid web development frameworks
combined with simple grid support libraries can be used to produce
useful science gateways.

\section{Future Work}

Although AMP is currently being used for friendly user testing and we do
not anticipate making any fundamental changes over the next year or two,
we have identified several front-end and back-end features that we wish
to explore in the future. Again, we are currently investigating the best
way to provide simulation progress and star result updates via RSS and
refining our use of AJAX techniques to enhance the user experience in
subtle yet meaningful ways. As the number of simulations on AMP grows,
we anticipate that we will need to revisit the interface used to
organize and present the results of the simulations.

One limitation of GridAMP that we intend to examine in the near future
is its use of multiple sequential GRAM jobs to propagate optimization
runs to completion. Although each GRAM job is set to the target system's
walltime (usually 6 or 24 hours), continuation jobs are only submitted
once the prior job has finished. Thus, the continuation jobs must wait
in the remote system's batch queue before processing can resume. Many
schedulers in use at TeraGrid sites support job chaining (or job
dependencies) such that multiple jobs can be submitted at once and
queued independently but declared elegible to run only after a prior job
has completed. This would be perfect for AMP jobs, as the initial
simulation submission could include the 4-8 jobs that are always
required to perform the simulation, possibly reducing the cumulative
queue wait time. We are currently making a graphical tool that plots job
wait vs. execution time on a Gantt chart for each AMP simulation, as
well as calculating aggregate execution wait and run time statistics, in
order to understand the impact of queue wait time on various systems. We
will then investigate Grid-based (but possibly nonstandard) methods to
submit chained jobs on the resources at the providers that are the most
tolerant of AMP's computational workloads.

\section{Conclusions}

AMP has provided an opportunity to develop a new science gateway
targeting TeraGrid computational resources. AMP's straightforward
workflow provided an ideal project to explore the use of the
Python-based Django web framework for rapid prototyping and development
of a science gateway. Our separation of the web interface, processing
daemon, and science components simplified the system's architecture and
implementation. Furthermore, our use of common Django modules for both
the web interface and the workflow daemon greatly reduced the complexity
of implementation. The entire workflow was easily implemented using
manual Globus command-line client calls to remote scripts and
executables, further simplifying debugging and allowing AMP to be
configured on remote resources without resource provider intervention.
AMP is currently available for friendly user testing, and we anticipate
the first extensive use of the system to perform new asteroseismology
science using Kepler data in October 2009. In the future, we plan to
examine possible applications of AMP's architecture and underlying
technology choices to other NCAR science gateway projects.

\section{Acknowledgments}

We would like to thank to Nancy Wilkins-Diehr and the TeraGrid Gateways
Program for assistance turning AMP into a TeraGrid science gateway, Stu
Martin for assistance with Globus GRAM auditing, and Tom Scavo for
assistance with GridShib. Thanks to Margaret Murray for helping us 
test GridAMP on TACC resources and to Victor Hazlewood and Rick Mohr for
assistance with NICS resources. Paul Marshall performed the
initial compilation and run time evaluation of MPIKAIA on several
TeraGrid resources. Will Baird developed many prototype AMP components
and features including AMP's utilization of the SIMBAD \cite{simbad}
astronomical database. Michael Oberg prepared and manages the NCAR
TeraGrid Service Hosting Platform used to host AMP and GridAMP.

Funding to integrate AMP with TeraGrid resources was provided by the
TeraGrid Science Gateways program. Computational time at NCAR was
provided by NSF MRI Grants CNS-0421498, CNS-0420873, and
CNS-0420985; NSF sponsorship of the National Center for
Atmospheric Research; the University of Colorado; and a grant from the
IBM Shared University Research program.
\\ % TODO -- CHECK PAGINATION

%\bibliographystyle{abbrv}
%\bibliography{gce09-amp}  % sigproc.bib is the name of the Bibliography in this case

% ACM needs 'a single self-contained file'!
%

%\balancecolumns

\end{document}